\documentclass[preprint,12pt,authoryear]{elsarticle}
\usepackage[latin1]{inputenc}
\usepackage{amsmath}
\usepackage{amsfonts}
\usepackage{amssymb}
\usepackage{makeidx}
\usepackage{graphicx}
\usepackage{natbib}
\usepackage{epstopdf}
\usepackage{booktabs}
\usepackage{breqn}

\usepackage{array}
\usepackage{float}
\usepackage{multirow}
\usepackage{amsbsy}
\usepackage{amstext}
\usepackage{amsthm}
\usepackage{amssymb}
\usepackage{graphicx}
\usepackage{esint}
\usepackage{url}
\usepackage{epstopdf}
\usepackage{lineno,hyperref}
\usepackage{tikz}
\usepackage{pgfgantt}
\usepackage{pdflscape}

\newtheorem{defn}{Definition}
\newtheorem{prop}{Proposition}
\newtheorem{thm}{Theorem}

\newtheorem{cor}{Corollary}
\newtheorem{proper}{Property}
\newtheorem{asum}{Assumption}
\DeclareMathOperator*{\argmax}{arg\,max}

\journal{Applied Energy}

\begin{document}

\begin{frontmatter}{}

\title{Limiting gaming opportunities on incentive-based demand response programs}

\author[focal]{José~Vuelvas\fnref{fn1} \corref{ca}}
\cortext[ca]{Corresponding author}

\author[focal]{Fredy~Ruiz\fnref{fn2}}

\author[focal2]{Giambattista~Gruosso\fnref{fn3}}

\fntext[focal]{E-mail addresses: vuelvasj@javeriana.edu.co (J. Vuelvas), ruizf@javeriana.edu.co (F. Ruiz) and giambattista.gruosso@polimi.it (G. Gruosso).}

\address[focal]{Departamento de Electrónica, Pontificia Universidad Javeriana, Bogotá, Colombia}
\address[focal2]{Dipartimento di Elettronica, Informazione e Bioingegneria, Milano, Italy}

\begin{abstract}
     Demand Response (DR) is a program designed to match supply and demand by modifying consumption profile. Some of these programs are based on economic incentives, in which, a user is paid to reduce his energy requirements according to an estimated baseline. Literature review and practice  have shown that the counter-factual models of employing baselines are vulnerable for gaming. Classical solutions of mechanism design require that agents communicate their full types which result in greater difficulties for its practical implementation. In this paper, a novel contract is developed to induce individual rationality (voluntary participation) and asymptotic incentive-compatibility (truthfulness) through probability of call, where an agent does not require to report the marginal utility. In this approach, a consumer only announces the baseline and reduction capacity, given a payment scheme that includes cost of electricity, incentive price, and penalty caused by any deviation between self-reported and actual energy consumption. The aggregator decides randomly what users are called to perform the energy reduction. As result, asymptotic truth-telling behavior in incentive-based DR is managed by the aggregator through the probability of call for each agent. Mathematical proofs and numerical studies are provided to demonstrate the properties and advantages of this contract in limiting gaming opportunities and in terms of its implementation.                    
 \end{abstract}

\end{frontmatter}{}

\section{Introduction}

DR is one of the most vital parts of the
future smart grid \citep{Zhu2013,Bloustein2005,Su2009}. DR is a tool to improve the demand profile, that is, to control the noncritical loads \cite{Buber2013} at the
residential \cite{Bizzozero2016}, commercial and industrial levels for matching supply
and demand. There are different ways to active DR in the electric power. Broadly defined, controllable programs \citep{Diaz2017} and indirect methods \citep{Vuelvas2017} are found as DR solutions, which are tools implemented by system operator (SO) to balance the demand with power generation by means of load modification. In particular, indirect methods are performed by changing energy price or giving an incentive payment.\\
In incentive-based DR, participating agents are paid for diminishing their energy consumption from established baseline (e.g. Peak Time Rebates, Interruptible Capacity Program and Emergency DR). Literature and practice have exhibited that consumers have incentives to alter their consumption patterns and baseline setting in order to increase their well-being, see e.g.  \citep{Zhou2017, Muthirayan2016,SeverinBorenstein2014, Vuelvas2015,Vuelvas2017}.There are three key components of an incentive-based DR program: 1) A baseline, 2) A payment scheme and 3) Terms and conditions (such as penalties) \citep{Muthirayan2016}.\\
The baseline is defined as a method of estimating the power consumption that would have been consumed by demand in the absence of DR \cite{Deng2015}. This quantity is often based on the average historical consumption of a consumer or a customer group on days that are similar to the forthcoming DR event. Therefore,
a counter-factual model is developed to estimate the customer baseline. In \cite{Mohajeryami2016}, some methods are presented to estimate the customer baseline. Baseline model error associated with DR parameter estimates are studied in \citep{Mathieu2011}. In addition, the performance of DR baselines are studied and new methods are proposed to obtain a reasonable compensation for consumers in \citep{Faria2013,Wijaya2014,Antunes2013}. Furthermore, in \citep{Chao2011}, the critical facts on the selection of customer baseline are highlighted, showing that counter-factual forecasts are vulnerable for gaming and could result in illusory demand reduction, then author proposes a baseline focusing on administrative and contractual approaches in order to get an efficient DR. \\ 
Given the fact that the incentive-based DR presents gaming concerns, a mechanism design or a contract is required to address these problems in order that an agent reveals his truthful baseline and private information.  Some solutions for DR are found in the literature by designing programs in a market framework. A revelation mechanism is developed in \citep{Ramos2013}, which requests agents to select the best choice for themselves among a menu of incentives in electricity markets. In \citep{Bitar2013}, a forward market is proposed, where users permit a deferred service in exchange for a reduced price of energy in order to manage the variability in supply from renewable generation, as well, this mechanism shows that prices are incentive-compatible. In \citep{Mhanna2014}, a two-stage mechanism is devised to share the cost of electricity among participants based on their day-ahead allocations. This mechanism is asymptotically incentive-compatible and ex-ante weakly budget balanced under certain conditions. Furthermore, a methodology is developed in \citep{Haring2014} for implementing microeconomic theory on contract designs for ancillary services in which incentive compatibility and individually rational are guaranteed in the presence of imperfect information between the consumers and the aggregator. In addition, in \citep{Okajima2014}, two kinds of mechanisms are applied for a dynamic day-ahead market based on DR. One is the Vicrey-Clarke-Groves (VCG), which is ex-post incentive compatible and individually rational. Another mechanism is Arrow,  d'Aspremont and G$\mathrm{\acute{e}rard}$-Varet, which is Bayesian (interim) incentive compatible and budget balanced. \\
In addition, other solutions are found in literature at the level of consumer and aggregator. In \citep{Ma2017}, a truthful mechanism is designed that uses a reward-bidding approach where the mechanism
adopts a fixed penalty for non-response for all selected agents, and consumers are selected in increasing order of their minimum acceptable rewards given the penalty. 
A model of consumer behavior in response to incentives is designed in a mechanism design framework in \citep{Zhou2017}, where aggregator collects the price elasticities of the demand as bids and then it selects the most susceptible users to incentives such that an aggregate reduction is obtained. In \citep{Fahrioglu2001}, truthful contracts are designed  for DR, which maximize the power utility company benefit function subject to individual rationality and incentive-compatible constraints. A VCG mechanism applied in DR is presented in \citep{6266724}, where the authors verify some properties such as efficiency, user truthfulness, and nonnegative transfer. However, there are several major obstacles implementing VCG mechanisms, see e.g. \citep{Roughgarden2016}. Other work of VCG approach is found in \citep{Meir2017}.  The authors of \citep{Muthirayan2016} propose an incentive-based DR mechanism, where each consumer reports his baseline consumption and his marginal utility to the aggregator. Deviations are penalized, hence, the true estimation is found. A linear utility function is assumed for each user. Furthermore, it is shown individual rationality for every consumer. In \citep{Chen2012}, a game theoretic DR strategy is developed, which consists of a distributed load prediction system by the participation of users that guarantee cheat-proof (truth-telling) behavior.  Finally, in \citep{Dobakhshari2016, Dobakshari2016}, a contract between a customer  and an aggregator for incentive-based DR is proposed. This mechanism is composed of two parts: a share of aggregator profit and a compensation paid to customer due to load reduction.\\
In this paper, a novel contract is addressed by the probability of call for each participant consumer. A user submits his baseline and reduction capacity. This contract does not require marginal utility information as in traditional mechanisms, which could be a private parameter difficult to estimate by a consumer. Accordingly, agents bid two quantities in terms of energy, then this contract is a more intuitive procedure and can be suitable to be implemented in practice. In this approach, the main goal of the aggregator is to select randomly which participant agents are called to performed DR. The contribution is summarized as follows:
\begin{itemize}
	\item The optimal decision problem is presented by 2-stage contract. The result is obtained backward in time to find optimal choice that user faces at each time. Theoretical analysis and numerical studies are provided to demonstrate the benefits and properties. The outcome shows that the contract is individually rational, incentive compatibility on the reported reduction capacity and asymptotic incentive compatibility on the reported baseline.    
	\item In this contract, aggregator does no require to estimate the customer baseline, then it only calls randomly participant consumers with a probability of call close to zero in order to obtain truthful behavior by demand side. In addition, There is no need for agents to inform their full types to achieve good performance in properties of this mechanism. 
\end{itemize}
The remainder of the paper is organized as follows. Section II describes the preliminary setting.  In Section III, the problem statement is explained. In Section IV, the proposed contract is formulated. In Section V, consumer's optimal choices are developed. In Section VI, Numerical optimization results are shown. In Section VII, a final discussion is presented. Conclusions are presented in Section VIII.

\section{Setting}
This part presents definitions, assumptions and preliminary considerations. Incentive-based DR program is managed by an aggregator that requests customers to curtail demand in response to an economic incentive. There are $I$ participant consumers. The set of users is denoted by $\mathcal{I}=\{1, 2, ..., I\}$. Some concepts and conditions are defined as follows.\\ 
The decision maker's preferences are specified by giving a smooth utility function $G(q_{i};b_{i})$, that depicts the level of satisfaction obtained by a user
as a function of  $q_{i}$,  which is the energy consumption and $b_{i}$ is the baseline, which is a quantity only known by each consumer $i\in\mathcal{I}$. The utility function
satisfies the following properties as proposed in \cite{Chen2012,Vuelvas2017,Vuelvas2015}:

\begin{proper}
	$G(q_{i};b_{i})$ is assumed as a
	concave function with respect to $q_{i}$. This implies that the
	marginal benefit of users is a nonincreasing function, i.e., $\frac{d^{2}G(q_{i};b_{i})}{dq_{i}^{2}}\leq0$.
	
\end{proper}

\begin{proper}
	Utility function is nondecreasing. Therefore, the marginal
	benefit is nonnegative $\frac{dG(q_{i};b_{i})}{dq_{i}}\geq0$.
\end{proper}

\begin{proper}
	$G(q_{i};b_{i})$ is zero when the consumption level is zero, $G(0)=0$.
\end{proper}
The energy price $p$ is given. Then, the following definition are stated.

\begin{defn} \label{def1}
	The energy total cost is $\pi_{i}(q_{i})=pq_{i}$. 
\end{defn}

\begin{defn} \label{def2}
	The payoff function without DR is defined as $U_{i}(q_{i};b_{i})=G(q_{i};b_{i})-\pi_{i}(q_{i})$,
	which indicates the user benefit of consuming $q_{i}$ for a certain time.
\end{defn}

\begin{defn}  \label{def:normal}
	The rational behavior
	of a consumer that maximizes the payoff function $U_{i}(q_{i};b_{i})$
	is given by, $q_{i}^{*}=b_{i}$. This result is found by solving the optimization problem $q_{i}^{*}=\arg\max_{q_{i}\in\{0, q_{max,i}\}}\:U_{i}(q_{i};b_{i})$. Where $q_{max,i}$ is the maximum allowable consumption value by user $i$.	
\end{defn}

\begin{asum} \label{defG}
	Under previous properties, the utility function can be approximated by a second order polynomial. User's utility function is assumed as,
	\[
	G(q_{i};b_{i})=\left\{ \begin{array}{cc}
	-\frac{\gamma_{i}}{2}q_{i}^{2}+[\gamma_{i}b_{i}+p]q_{i} & \quad0\leq q_{i}\leq b_{i}+\frac{p}{\gamma_{i}}\\
	\frac{p^{2}}{2\gamma_{i}}+\frac{\gamma_{i}b_{i}^{2}}{2}+pb_{i} & \quad q_{i}>b_{i}+\frac{p}{\gamma_{i}}
	\end{array}\right.
	\]
	where $\gamma_{i}$ is the marginal utility of consumer $i$.		
	
\end{asum}

Therefore, given Assumption \ref{defG} and Definition \ref{def:normal}, the optimal payoff $U_{i}(q_{i};b_{i})$ for an agent $i$ that does not participate in DR is equal to 

\begin{equation}\label{payoffnoDR}
\frac{\gamma_{i}b_{i}^2}{2} 
\end{equation}

\begin{asum}
	Each consumer $i$ has a maximum limit of energy consumption $q_{max,i}$. A reasonable hypothesis is that the maximum limit is greater than the saturation limit established in Assumption \ref{defG}, i.e., $q_{max,i}>b_{i}+p/\gamma_{i}$
\end{asum}

\section{Problem statement}
\begin{figure}[ht]
	\begin{center}
		\includegraphics[scale=1]{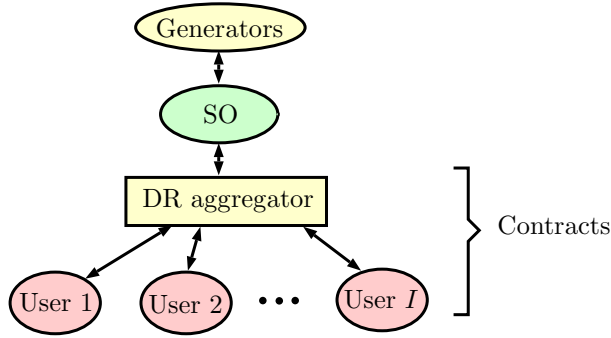}
	\end{center}
	\caption{\label{fig:esq_aggregator} Participant agents.}
\end{figure}

In a wholesale electricity market, according to the available energy sources, aggregated demand information and market prices, SO could require a demand reduction during peak times. Therefore, SO request a DR process through an aggregator. Fig. \ref{fig:esq_aggregator} shows the participant agents involved in DR. Aggregator can participate in the market by bidding aggregated reduction capacity to the system. After the market clearing, SO sends a demand reduction requirement to aggregator. Then, by means of contracts with consumers, aggregator activates DR programs. 

In an incentive-based DR approach, aggregator encourages to each participant diminishing the energy pattern according to an economic signal.  Mathematically,       

\begin{defn} \label{def8}
	Let $p_{2}$ be the rebate price received by the user due to energy reduction
	in peak periods. The incentive is defined as $\pi_{i,2}(q_{i};\hat{b_{i}})=p_{2}(\hat{b_{i}}-q_{i})_{+}$. Where $(\cdot)_{+}=\max(\cdot,0)$ and the superscript $\ \hat{} $ means declared information. Therefore, $\hat{b_{i}}$ is the baseline announced by user. 
\end{defn}

\begin{asum}
	In order to encourage the energy cutback by demand side. The incentive price is assumed greater or equal than energy price, that is, $p_{2}\geq p$.
\end{asum}

In order to obtain an efficient DR service, aggregator requires that each agent reports his true private information (or estimated data on non-altered past consumption) and selects his consumption level according to his preferences, i.e.: 
\begin{equation}
q_{i}^{o}=\argmax_{q_{i}\in\{0, q_{max,i}\}}\:U_{i}(q_{i};b_{i})-\pi_{i}(q_{i})+r\pi_{i,2}(q_{i};\hat{b_{i}}) \label{idealDR}
\end{equation}
with $\hat{b_{i}}=b_{i}$ and $r_{i}$ a binary variable that indicates if user $i$ is called to participate in DR. In other words, truth-telling behavior is desired when incentive-based DR is required during peak times of electrical demand. Under assumption that consumers behave as truthful agents, consumer's optimal choice is given by the following proposition.

\begin{prop} \label{prop1}
	The optimal consumption $q_{i}^{o}$ of a user that participates in an incentive-based DR, given truthful report ($\hat{b_{i}}=b_{i}$) is:
	\[
	q_{i}^{o}=\begin{cases}
	b_{i} & r_{i}=0\\
	(b_{i}-\frac{p_{2}}{\gamma_{i}})_{+} & r_{i}=1
	\end{cases}
	\]
\end{prop}   	

Case $r_{i}=0$ follows from Definition \ref{def:normal}. Case $r_{i}=1$ is derived from optimality conditions of problem (\ref{idealDR}) at each interval of $G(q_{i};b_{i})$, and selecting the global maximizer (see \cite{Vuelvas2017}).

Notice that Prop. \ref{prop1} is the consumer's ideal behavior. However, in practical fashion, a user can alter his reported baseline $\hat{b_{i}}$, in particular, he could bid $q_{max,i}$ because there is no penalty to increase his well-being. Furthermore, in DR programs like Peak Time Rebate where the baseline is estimated, a consumer modifies as well the average consumption at similar previous time resulting in the problem described in \cite{SeverinBorenstein2014}, for the case when $p_2>p$, he consumes up to $q_{max,i}$ during the baseline setting period  (see \cite{Vuelvas2017,Vuelvas2015}). Therefore, in this paper, a contract between the aggregator and each consumer is proposed to induce asymptotic incentive compatible (truthfulness) and individually rational (voluntary participation) properties.

\section{Contract based on probability of call}

\begin{figure}[ht]
	\begin{center}
		\includegraphics[scale=1]{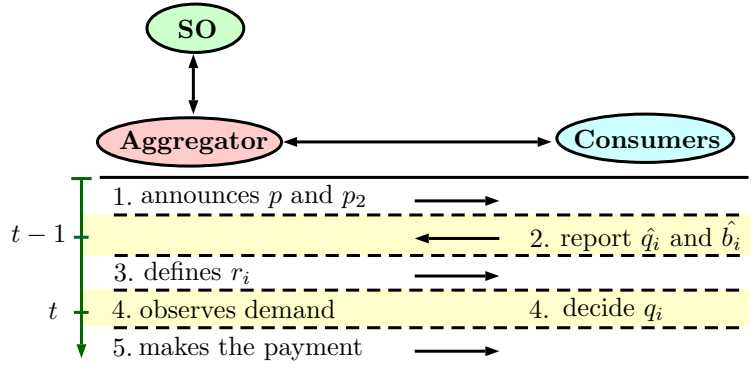}
	\end{center}
	\caption{\label{fig:contract} Timeline of the contract.}
\end{figure}

In this section, the proposed contract is described according to a new payment scheme based on probability of call. Fig. \ref{fig:contract} shows the timeline to hold the contract between the aggregator and an agent. The agreement process is described below. 
\begin{itemize}
	\item First, SO informs to aggregator what net demand reduction is required at a certain peak time or the aggregator could participate as new player within electricity market. Then aggregator announces energy cost $p$ and incentive price $p_{2}$ (Also, this quantity works as penalty for deviation).
	\item  According to all prices, at the time $t-1$, participant consumers (that signed the contract) report the baseline $\hat{b_{i}}$ and their reduced energy consumption $\hat{q_{i}}$ if they are called.
	\item  Next, aggregator determines the subset of agents that are called to participate in DR subject to SO requirement and probability of call for each participant, i.e., aggregator decides the binary variable $r_{i}$.
	\item Later, at the time $t$, each user receives his value of $r_{i}$ and chooses his actual consumption level $q_{i}$ during the event. This consumption is observed by the aggregator.
	\item Finally, the payment is made by the aggregator for called users.
\end{itemize}

The proposed payment scheme of the contract is presented as follows:  

\begin{defn} \label{ps}
	Let $\pi_{i,3}^{r_{i}}(q_{i};\hat{b_{i}},\hat{q_{i}})$ be the aggregator payment scheme under incentive-based DR:
	\[
	\pi_{i,3}^{r_{i}}(q_{i};\hat{b_{i}},\hat{q_{i}})=\begin{cases}
	p\max(\hat{b_{i}},\,q_{i}) & r_{i}=0\\
	pq_{i}-p_{2}(\hat{b_{i}}-q_{i})_{+}+p_{2}\left|q_{i}-\hat{q_{i}}\right| & r_{i}=1
	\end{cases}
	\]   
	
\end{defn}
According to Definition \ref{ps}, note that if a consumer is not called to participate in DR, then he must pay the maximum between his reported baseline or actual energy consumption. This payment is inspired by the concept of "buy the baseline" \cite{Chao2011}. Otherwise, if he is required to reduce his energy requirement, then he receives an incentive of $p_2(\hat{b_{i}}-q_{i})$ and must pay the energy cost and a penalty with the same price of the incentive if his actual consumption differs from $\hat{q_{i}}$. Given the payment scheme established in Definition \ref{ps}, thus the contract is settled as a 2-stage procedure:\\ 
\textbf{Stage 1)} Given the prices $p$ and $p_{2}$, each consumer reports $\hat{b_{i}}$ and $\hat{q_{i}}$ to aggregator. \\  
\textbf{Stage 2)} Aggregator determines which users are called to participate in DR by means of the variable $r_{i}$. Each agent decides his actual energy consumption $q_{i}$ according to the aggregator decision $r_{i}$.\\
Finally, an important and logical rule of this contract is stated in the following hypothesis. 
\begin{asum}
	 The reported baseline must be greater or equal than reported energy consumption under DR, i.e, $\hat{b_{i}}\geq \hat{q_{i}}$.\label{asum:logical}
\end{asum}
In the next Section, consumer's optimal decision under this proposed contract is determined by mathematical proofs. These results are vital for designing the aggregator role.     

\section{Consumer's optimal choices}
For the previous DR contract, a consumer faces the problem of deciding what to communicate to aggregator and how much energy $q_{i}$ to consume. Given $p$ and $p_{2}$, a rational user finds $\hat{b_{i}}$, $\hat{q_{i}}$  and $q_{i}$ such that his profits $J_i$ will be maximized:    
\begin{equation}
[\hat{b_{i}}^{*},\hat{q_{i}}^{*},q_{i}^{*}]=\argmax_{q_{i},\hat{b_{i}},\hat{q_{i}}\in\{0, q_{max,i}\}}\:\, J_i=\mathbf{E}(G(q_{i};b_{i})-\pi_{i,3}^{r_{i}}(q_{i};\hat{b_{i}},\hat{q_{i}})) \label{cp}   
\end{equation}
The superscript $^{*}$ means optimal decisions.\\
Problem (\ref{cp}) is a two-stage stochastic programming formulation. The proposed solution is found by solving backward in time \citep{Rajagopal2013}.\\
At the time $t$ (see Fig. \ref{fig:contract}), user maximizes the actual energy consumption $q_{i}$ for each case of $r_{i}$, that is, finds the optimal solution $q_{i,r_{i}=0}^{*}$ and $q_{i,r_{i}=1}^{*}$, given $\hat{b_{i}}$ and $\hat{q_{i}}$. 
\begin{equation}
[q_{i,r_{i}=0}^{*},q_{i,r_{i}=1}^{*}]=\argmax_{q_{i,r_{i}}\in\{0, q_{max,i}\}}\:\, G(q_{i,r_{i}};b_{i})-\pi_{i,3}^{r_{i}}(q_{i,r_{i}};\hat{b_{i}},\hat{q_{i}}) \label{stage2}
\end{equation}
At the time $t-1$, consumer determines the best information $\hat{b_{i}}^{*}$ and $\hat{q_{i}}^{*}$ to report, knowing the optimal choices $q_{i,r_{i}=0}^{*}$ and $q_{i,r_{i}=1}^{*}$ that he can take at consumption time $t$ and facing the uncertainty in $r_{i}$.  
\begin{dmath}
	[\hat{b_{i}}^{*},\hat{q_{i}}^{*}]=\argmax_{\hat{b_{i}},\hat{q_{i}}\in\{0, q_{max,i}\}}\:\, \mathbf{E}(G(q_{i,r_{i}}^{*};b_{i})-\pi_{i,3}^{r_{i}}(q_{i,r_{i}}^{*};\hat{b_{i}},\hat{q_{i}}))=\argmax_{\hat{b_{i}},\hat{q_{i}}\in\{0, q_{max,i}\}}\:\, p_{r_{i}}[G(q_{i,r_{i}=1}^{*};b_{i})-\pi_{i,3}^{r_{i}=1}(q_{i,r_{i}=1}^{*};\hat{b_{i}},\hat{q_{i}})]+[1-p_{r_{i}}][G(q_{i,r_{i}=0}^{*};b_{i})-\pi_{i,3}^{r_{i}=0}(q_{i,r_{i}=0}^{*};\hat{b_{i}},\hat{q_{i}})]  \label{stage1}
\end{dmath}
where $p_{r_{i}}=p_{r_{i}}(r_{i}=1)$ is the probability of call.\\ 
The optimal decision problem is described in details as follows.  
\subsection{Second-stage of consumer's decision}
The problem (\ref{stage2}) is solved for each case of the binary variable $r_{i}$. The solutions are presented in Theorems \ref{thr0} and \ref{thr1}. The proofs are shown in \ref{ath0} and \ref{ath1}, respectively. 
 
\begin{thm}
	The optimal consumption $q_{i,r_i}^{*}$ for the signal $r_i=0$ of a participant consumer in the proposed contract, given the problem (\ref{stage2}), is:
	\[
	q_{i,r_{i}=0}^{*}=\begin{cases}
	b_{i}& 0 \leq \hat{b_{i}} \leq b_{i} \qquad strategy\,\mathcal{A}\\
	\hat{b_{i}}& b_{i} < \hat{b_{i}} \leq b_{i}+p/\gamma_{i}\qquad strategy\,\mathcal{B}\\
	b_{i}+p/\gamma_{i} & b_{i}+p/\gamma_{i} < \hat{b_{i}} \leq q_{max,i}\qquad strategy\,\mathcal{C}  
	\end{cases}
	\]
	\label{thr0}
\end{thm} 

According to Theorem \ref{thr0}, there are three strategies that depend on the reported baseline. For instance, if a consumer informs a baseline above his preferences and below his saturation limits (i.e. $strategy\,\mathcal{B}$), then, his best choice is to consume what he reported in the previous stage.

\begin{thm}
	The optimal consumption $q_{i,r_i}^{*}$ for the signal $r_i=1$ of a participant consumer in the proposed contract, given the problem (\ref{stage2}), is:
{\footnotesize 	\[
	q_{i,r_{i}=1}^{*}=\begin{cases}
	b_i&b_i\leq \hat{b_{i}} \leq q_{max,i}, b_i \leq \hat{q_{i}} \leq \hat{b_{i}} \leq q_{max,i} \qquad strategy\,\mathcal{U}\\
	\hat{q_{i}}& b_i-\frac{p_2}{\gamma_{i}}\leq \hat{b_{i}} \leq q_{max,i}, b_i-\frac{2p_2}{\gamma_{i}}\leq \hat{q_{i}} \leq \hat{b_{i}} \leq b_i \qquad strategy\,\mathcal{V}\\
	\hat{q_{i}}&\alpha \leq \hat{b_{i}} \leq b_i-\frac{p_2}{\gamma_{i}}, b_i-\frac{2p_2}{\gamma_{i}} \leq \hat{q_{i}} \leq b_i-\frac{p_2}{\gamma_{i}}\qquad strategy\,\mathcal{W}\\
	(b_i-\frac{p_2}{\gamma_{i}})_{+}&b_i-\frac{2p_2}{\gamma_{i}}\leq \hat{b_{i}} \leq \alpha, b_i-\frac{2p_2}{\gamma_{i}}\leq \hat{q_{i}} \leq \hat{b_{i}} \leq b_i-\frac{p_2}{\gamma_{i}} \qquad strategy\,\mathcal{X}\\
	(b_i-\frac{p_2}{\gamma_{i}})_{+}&0\leq\hat{b_{i}}\leq b_i-\frac{3p_2}{2\gamma_{i}}, 0\leq \hat{q_{i}} \leq \hat{b_{i}} \leq b_i-\frac{2p_2}{\gamma_{i}}\qquad strategy\,\mathcal{Y}\\
	(b_i-\frac{2p_2}{\gamma_{i}})_{+}&b_i-\frac{3p_2}{2\gamma_{i}}\leq \hat{b_{i}} \leq q_{max,i}, 0 \leq \hat{q_{i}} \leq b_i-\frac{2p_2}{\gamma_{i}} \qquad strategy\,\mathcal{Z} 
	\end{cases}
	\]}
	with $\alpha=\frac{\gamma_{i}b_i^2}{2p_2}-\frac{\gamma_{i}b_i\hat{q_{i}}}{p_2}-b_i+\frac{\gamma_{i}\hat{q_{i}}^2}{2p_2}+2\hat{q_{i}}+\frac{p_2}{2\gamma_{i}}$. 
	\label{thr1}
\end{thm}

 Similarly, Theorem \ref{thr1} presents the consumer's optimal decisions conforming to $\hat{b_{i}}$ and $\hat{q_{i}}$. In this case, notice that the strategies that enable an energy reduction are $strategies\,\mathcal{X}$, $\mathcal{Y}$ and $\mathcal{Z}$. Next, these results are substituted in (\ref{stage1}) in order to find profit-maximizing behavior.

\subsection{First-stage of consumer's decision}
The solution of problem (\ref{stage1}) is written in Theorem \ref{threport}. The proof is described in \ref{athreport}.  

\begin{thm} 
	Given $q_{i,r_i}^{*}$ from Theorems \ref{thr0} and \ref{thr1}, then the optimal reports $\hat{b_{i}}^{*}$ and $\hat{q_{i}}^{*}$ for (\ref*{stage1}) are:  
	\[
	\hat{b_{i}}^{*}=\begin{cases}
	\frac{p_{r_i}p_2}{\gamma_{i}(1-p_{r_i})}+b_i & 0 \leq p_{r_i}  \leq \frac{p}{p_2+p} \\
	q_{max,i} & \frac{p}{p_2+p} \leq p_{r_i} \leq 1\\
	\end{cases}
	\]
	
	\[
	\hat{q_{i}}^{*}=\left(b_i-\frac{p_2}{\gamma_{i}}\right)_{+}
	\]
	\label{threport}
	
\end{thm}

Theorem \ref{threport} presents the optimal reported decisions for rational consumers in this contract. With respect to the reported baseline $\hat{b_{i}}^{*}$, if the probability of call $p_{r_i}$ is less than threshold $p/(p_2+p)$, then an agent informs his true baseline $b_i$ added to $p_2/\gamma_{i}$ which is multiplied by the expression $p_{r_i}/(1-p_{r_i})$. For instance, if $p_{r_i}$ is close to zero then the best choice is to announce $b_i$. Therefore, the term $p_{r_i}/(1-p_{r_i})$ limits the misreporting by means of $p_{r_i}$. Note that the probability of call is limiting gaming opportunities for the baseline under this agreement. Otherwise, when a user knows that his probability of call exceeds the threshold thus the best strategy is to report his maximum energy allowed $q_{max,i}$. 
In addition, a consumer must inform the energy reduction $\hat{q_{i}}$, in this affair, the truth information is obtained independently of the probability of call according to Theorem \ref{threport}.       

\begin{cor} \label{EprofitDR}
	The optimal expected profit $J^{*}$ is: 
\[
	J_i^{*}=\begin{cases}
	\frac{b_i^2\gamma_{i}}{2}+\frac{p_{r_{i}} p_2^{2}}{2\gamma_{i}(1-p_{r_{i}})}& 0 \leq p_{r_i}  \leq \frac{p}{p_2+p} \\
	\frac{p^2}{2\gamma_{i}}+b_ip-pq_{max,i}+\frac{b_i^2\gamma_{i}}{2}-\frac{p^2p_{r_{i}}}{2\gamma_{i}}+\frac{p_2^2p_{r_{i}}}{2\gamma_{i}}\\-b_ip_{r_{i}}-b_ip_2p_{r_{i}}+pq_{max,i}p_{r_{i}}+p_2q_{max,i}p_{r_{i}} & \frac{p}{p_2+p} \leq p_{r_i} \leq 1\\
\end{cases}
\]
	
\end{cor}

Corollary \ref{EprofitDR} is found by substituting solution of Theorem \ref{threport} in Eq. (\ref{cp}). It is easy to prove that the optimal expected profit $J^{*}$ is an increasing function w.r.t. $p_{r_i}$. Relevant cases are presented when the probability of call is lower than the threshold $p/(p_2+p)$. Notice that when $p_{r_i}$ goes to zero, the optimal expected profit under DR is $b_i^2\gamma_{i}/2$, which is the same value for a user that does not participate in DR as indicated by  (\ref{payoffnoDR}).       

\begin{cor}
	The optimal consumption $q_{i,r_i}^{*}$, by replacing the best reports given in Theorem \ref{threport}, is:
	\[
	q_{i,r_i}^{*}=\begin{cases}
	\frac{p_{r_i}p_2}{\gamma_{i}(1-p_{r_i})}+b_i & r_i=0, \, 0 \leq p_{r_i}  \leq \frac{p}{p_2+p} \\
	b_{i}+p/\gamma_{i} & r_i=0, \, \frac{p}{p_2+p} \leq p_{r_i} \leq 1\\
	(b_i-\frac{p_2}{\gamma_{i}})_{+} & r_i=1, \, 0 \leq p_{r_i} \leq 1 \\
	\end{cases}
	\] \label{cor:ree}
\end{cor}

Lastly, the optimal decision $q_{i,r_i}^{*}$ is determined by using Theorem \ref{threport} in the second-stage defined in Eq. \ref{stage2}. On the one hand, if an agent is not called to participate in DR and the probability of call is below the threshold, then he should consume the energy he reported in order to earn his best profit. However, if $p_{r_i}$ is above of the threshold $p/(p_2+p)$ when $r_{i}=0$, so a rational consumer uses energy to saturation point which is $b_{i}+p/\gamma_{i}$. On the other hand, if a user is called to participate $r_{i}=1$, then his best choice is to consume what he reported in the previous stage.

\subsection{Contract properties}
Finally, contract properties are direct consequences of previous results. These features are listed below. 

\begin{cor} \label{cor:rational}
	Individually rational (voluntary participation): a user that participates in this approach obtains a profit at least as good as he does not sign the DR contract.
\end{cor}

An important requirement to encourage affiliation to the contract or mechanism is the property stated in Collorary \ref{cor:rational}. This results follows by comparing Corollary \ref{EprofitDR} and expression (\ref{payoffnoDR}). For rational users, other income can be incorporated into their economic activities by participating in proposed settlement.

\begin{cor} \label{cor:reportedconsumption}
	Incentive compatibility on the reported energy consumption under DR: a consumer informs the truthful consumption under DR according to his preferences.  
\end{cor}

\begin{cor} \label{cor:baseline}  
	Asymptotic incentive compatibility on the reported baseline: as the probability of call tends to zero, the consumer's optimal strategy is to declare $\hat{b_{i}}=b_i$.  
\end{cor}

Corollaries \ref{cor:reportedconsumption} and \ref{cor:baseline} establish the truthful properties in this contract. These features are understood by reviewing Theorem \ref{threport} and Corollary \ref{cor:ree}.  A rational user bids $\hat{q_{i}}^{*}=q_{i,r_i=1}^{*}$ guarantying to maximize his profit.  Additionally, if $p_{r_i}\longrightarrow0$ then the best choice is to announce $\hat{b_{i}}^{*}=q_{i,r_i=1}^{*}$, i.e., asymptotic truthfulness. 

\section{Illustrative example}

In this section, simulation results are presented to illustrate the
optimal behavior of a user when he is participating in the proposed contract. The retail price is $p= 0.26$ $\$$/kWh (based on peak summer rate in 10/1/16 by Pacific Gas and Electric Company in San Francisco, California), true baseline is $b_{i}=8$ kWh, the incentive/penalty price is $p_{2}=0.3$ $\$$/kWh, the marginal utility is $\gamma_{i}=0.05$ $\$$/kWh$^2$ and the maximum allowable consumption is $q_{max,i}=16$ kWh. Randomness $r_{i}$ has been created to simulate consumer's optimal choice according to the probability of call. A Monte Carlo simulation is performed with
$1000$ realizations of $r_{i}$ for each value of probability in order to check the results.

\begin{figure}[ht]
	\begin{center}
		\includegraphics[scale=0.6]{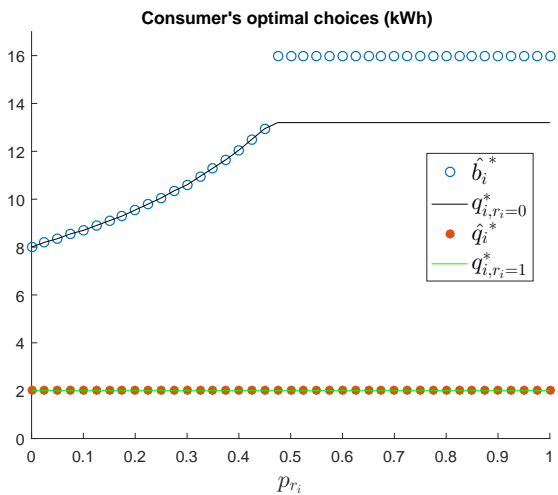}
	\end{center}
	\caption{\label{fig:optimal_choice} Optimal consumer's choice.}
\end{figure}

The problem (\ref{cp}) is solved through extensive simulations for different values of probability $p_{r_{i}}$ to find rational decisions. The consumer's optimal choices are depicted in Fig. \ref{fig:optimal_choice}. As regards the reported baseline, the truthfulness is being lost as the probability of call increases which is described by Theorem \ref{threport}. For this example, the threshold $p/(p+p_2)$ in the probability of call is $0.46$. Above the threshold, the worst condition of gaming are found because a user announces his maximum energy allowed and in order to achieve the highest income, as well, he really consumes until his energy saturation assuming that he is not called to participate in DR. Additionally, note that the agent actually declares what he is willing to reduce according to his preferences $\hat{q_{i}}^{*}=q_{i,r_i=1}^{*}$ irrespective of the probability of call because he suffers a penalty for deviations of his self-reported consumption when he is called to DR program. 

\begin{figure}[ht]
	\begin{center}
		\includegraphics[scale=0.6]{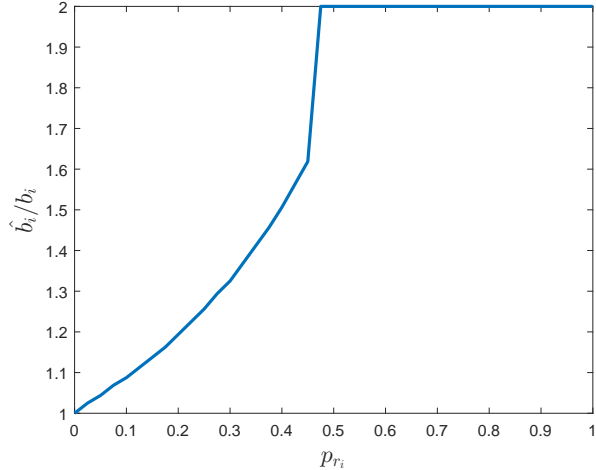}
	\end{center}
	\caption{\label{fig:percentage} Percentage of gaming limitation on the reported baseline.}
\end{figure}
Furthermore, Fig. \ref{fig:percentage} depicts the reported baseline normalized by the true one in order to establish the percentage of limitation on gaming opportunities. For instance, an aggregator calls a group of agents with a probability of call equals to $p_{r_i}=0.1$ then rational consumers have incentives to overreport the baseline through the increase of 11\% due to the expression $p_2p_{r_i}/(\gamma_{i}(1-p_{r_i}))$ given by Theorem \ref{threport} which is added to his true baseline. Accordingly, an aggregator does not know the values of $\gamma_{i}$ for all agents $i$, however, it can limit their gaming behavior by means of the probability of call.     

The intuition behind these findings is that, on the one hand, whether the participant user is not called then he must pay the maximum between his reported baseline and actual consumption, thus, misreporting will cause a loss of his profit. On the other hand, if the consumer is called, hence he should reduce his energy consumption according to what he committed at previous Stage so that the penalty does not apply. Moreover,  consumer's optimal decisions are coherent with Prop. \ref{prop1} given a probability of call close to zero. Therefore, asymptotic incentive compatibility is induced by means of this contract.

\begin{figure}[ht]
	\begin{center}
		\includegraphics[scale=0.6]{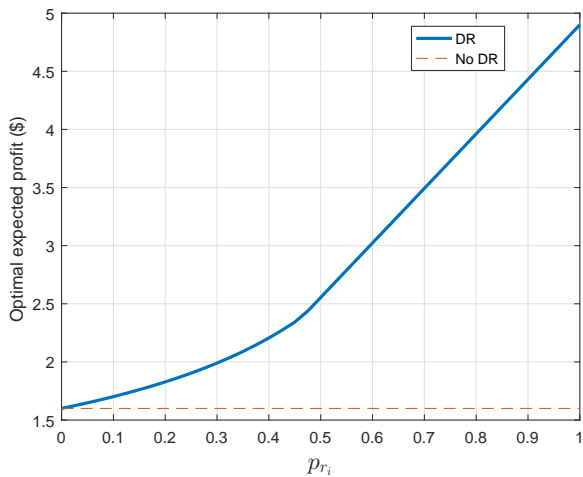}
	\end{center}
	\caption{\label{fig:expected} Optimal expected profit of a consumer.}
\end{figure}

Fig. \ref{fig:expected} shows the optimal expected profit of a consumer according to the probability of call. This curve is increasing with $p_{r_{i}}$ and two parts are distinguished as is stated in Collorary \ref{EprofitDR}. The first one, a smooth growth is exhibited for $p_{r_{i}}\in[0,0.46]$ then a significant rise in benefits is presented for $p_{r_{i}}>0.46$ as the second part. The point of change between the two parts is presented in the threshold probability of call, which is coherent with Fig. \ref{fig:optimal_choice}.  Furthermore, the payoff when an agent does not participate in DR is given by expression (\ref{payoffnoDR}) that for this case is \$1.6. In Fig. \ref{fig:expected}, the lowest optimal expected profit occurs in $p_{r_{i}}=0$ with a payoff of \$1.6, hence, the user's benefit when he participates in this contract is at least as good as when he does not join in the incentive-based DR program. Lastly, voluntary participation, also known as individual rationality, is motivated in the proposed contract.     

\begin{figure}[ht]
	\begin{center}
		\includegraphics[scale=0.6]{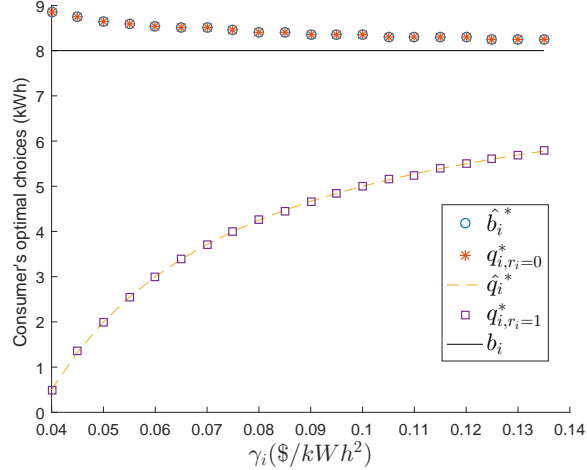}
	\end{center}
	\caption{\label{fig:gamma} consumer's optimal choice and optimal profit by varying private preference $\gamma_{i}$ given the probability of call $p_{r_{i}}=0.1$.}
\end{figure}    

Finally, variations in the preference $\gamma_{i}$ are analyzed in this contract. In Fig. \ref{fig:gamma}, the marginal utility $\gamma_{i}$ is changed given fixed probability of call and prices, in order to study the effects on the fluctuation of this consumer's private preference. For $p_{r_{i}}=0.1$, the optimal reported baseline $\hat{b_{i}}^{*}$ is equal to the actual energy consumption $q_{i,r_i=0}^{*}$ if he is not called for DR and also, as $\gamma_{i}$ increases, the reported baseline tends to the true baseline $b_{i}$. In addition, reported and actual energy consumption are the same $\hat{q_{i}}^{*}=q_{i,1}^{*}$, which is consistent to Theorem \ref{threport} and Corollary \ref{cor:ree}. In summary, as electrical energy preference $\gamma_{i}$ rises for a probability of call beneath of threshold value, overbidding in the baseline and reduction capacity diminish; also, the reported information and the actual consumption values hold the same behavior.

\begin{figure}[ht]
	\begin{center}
		\includegraphics[scale=0.6]{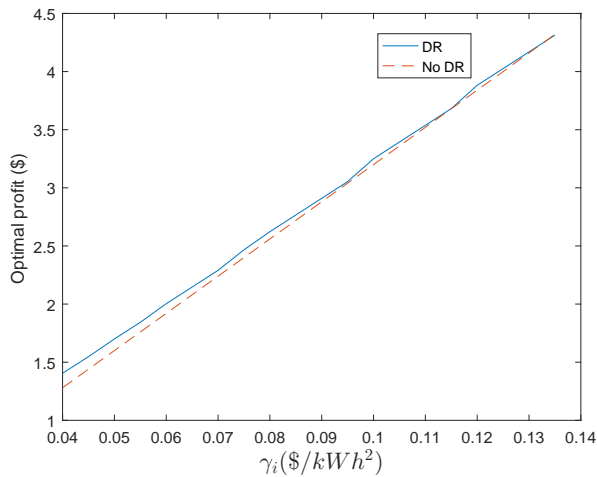}
	\end{center}
	\caption{\label{fig:gamma2} Optimal profit by varying private preference $\gamma_{i}$ given the probability of call $p_{r_{i}}=0.1$.}
\end{figure} 

Lastly, Fig. \ref{fig:gamma2} presents the optimal profit by changing $\gamma_{i}$. The curve trend is an increasing function with respect to $\gamma_{i}$ due to preliminary assumptions of the consumer behavior. Note that the property of voluntary participation is guaranteed irrespective of the consumer's preferences since a rational user (profit-maximizing) obtains greater economic profit by joining in DR rather than not participating.   

\section{Discussion}
In practical fashion, an industrial, commercial or residential consumer is not thinking in terms of marginal utility when he is dealing with DR. Most of the solutions found in literature, the agreement or mechanism requires that agents reveal all their types including the private value $\gamma_{i}$ in order to settle the problem. It is most intuitively that a user with smart metering devices can know or estimate how much energy he consumes during a certain hour of the day, thus he could announce information to the aggregator in terms of $kWh$. For instance, if an industrial consumer is willing to turn off some machines during DR event then he can determine his possible reduction of energy through knowledge of the equipment power. Therefore, a user with the suitable technology that analyzes the behavior of each appliance can take his decisions and participate in incentive-based DR programs by knowing his preferences of energy consumption. One of the advantages of the proposed contract is actually that the required information  of participants is in energy units. Although this new approach includes the analysis of marginal utility, the contract can be implemented without this knowledge.    

Traditional incentive-based DR programs rely on contractual models to estimate baseline in order to make the payments to participant users. This approach is vulnerable for gaming since consumers increase their energy consumption during baseline setting period harming the system reliability. The proposed contract induces an asymptotic incentive compatible property on the self-reported baseline by participant agents since the marginal utility is not available to settle the contract. In spite of this agreement feature, a regular consumer is not taking decisions with stochastic optimization tools, therefore, whether the aggregator keeps the probability of call close enough to zero then the truth-telling behavior is obtained under this novel approach.      

According to Corollary \ref{cor:reportedconsumption}, incentive compatibility on the reported energy consumption under DR is guaranteed due to the form of the penalty function which is given by a expression with absolute value, i.e., $p_{2}\left|q_{i}-\hat{q_{i}}\right|$. At first sight, it might seem that this penalty is very strict. Although, it is possible to relax this condition through tolerance or hysteresis band in order to incorporate some flexibility in the contract and that agents do not feel so rigid when consuming energy at Stage 2 if they are called to participate in DR. The final goal is to design a contract that could be implementable and elicit truthful properties for improving part of the power management in a smart grid.

\section{Conclusions}

In this paper was proposed a DR contract that induces voluntary participation and asymptotic truthful properties based on the probability of call for users. The problem was addressed using a two-stage stochastic programming algorithm. The formulation allowed linking the consumer's decisions between the different stages in order to determine the consumer's optimal behavior. It was found that if an aggregator keeps the probability of call close to zero then consumers reveal their truth information about energy preferences. Hence, the main goal of the aggregator is to call randomly a subset of users that meets the probability criterion and satisfies the energy reduction requirement demand by SO.  

A contract for incentive-based DR based on probability of call enables to limit the gaming opportunities by making that a consumer buys his self-reported baseline or pays his actual consumption if this one is greater than his bid when he is not called to participate in DR. The user's uncertainty under this agreement motivates him to inform the truth of his preferences. In addition, the model provides to the aggregator the limit allowed probability to maintain a bounded deviation in user behavior. Therefore, the limitation of gaming occurs beneath of threshold and its percentage is controlled by the probability of call.

For future directions, a model to control the operation of an aggregator that manages a portfolio of some incentive-based DR contracts which includes indirect and direct control programs would be a significant improvement for power management of a smart grid and assess the scope of the proposed contract.  Moreover, an important discussion would be to determine what kind of technology is required for all agents to implement and ensure system reliability in this agreement.           

\section*{Acknowledgements}
J. Vuelvas received a doctoral scholarship from COLCIENCIAS (Call 647-2014). Also, this work has been partially supported by COLCIENCIAS (Grant 1203-669-4538, Acceso Universal a la Electricidad). 

\appendix

\section{Proof of the Theorem \ref{thr0}}\label{ath0}

\begin{proof}
The optimization problem is analyzed by cases due to the non-linearities in the functions. Then several solutions are found according to the reported information given by a user at Stage 1. Finally, results are compared and then the global solution is determined. Considering the signal $r_i=0$, next, the optimization problem is posed as follows: 	
	
	\[
	q_{i,r_{i}=0}^{*}=\argmax_{q_{i,r_{i}=0}\in\{0, q_{max,i}\}}\:\, H_{r_{i}=0}=G(q_{i,r_{i}};b_{i})-p\max(\hat{b_{i}},\,q_{i})
	\]
Four cases are identified as results of payment function which is the maximum between reported baseline and actual energy consumption, and the two parts of utility functions $G(.)$ (see Def. \ref{defG}).  Cases are denoted by lowercase letters. A optimization problem is formulated for each case according to its conditions. 
	
The case a) is assumed when reported baseline is greater than the actual consumption, and the utility function is non-saturated, i.e.,
	
a) $r_{i}=0$, $\hat{b_{i}}\geq q_{i}$, $0\leq q_{i}\leq b_{i}+\frac{p}{\gamma_{i}}$, $0 \leq \hat{b_{i}} \leq q_{max,i}$
	\begin{alignat*}{3}
	&q_{i,r_{i}=0}^{*}=\quad&& \argmax_{q_{i,r_{i}=0}} \quad && H_{r_{i}=0}=-\frac{\gamma_{i}}{2}q_{i}^{2}+[\gamma_{i}b_{i}+p]q_{i}-p\hat{b_{i}}\\
	& && \text{s.t.}&& -\hat{b_{i}}+q_{i}\leq 0 \\
	& && && q_{i}-b_{i}-p/\gamma_{i} \leq 0 \\
	& && && -q_{i} \leq 0 
	\end{alignat*}
	The first-order optimality condition yields to
	\[
	q_{i,r_{i}=0}^{*}=\begin{cases}
	\hat{b_{i}} & 0 \leq \hat{b_{i}} \leq b_{i}+p/\gamma_{i}\\
	b_{i}+p/\gamma_{i} & b_{i}+p/\gamma_{i} \leq \hat{b_{i}} \leq q_{max,i}  
	\end{cases}
	\]
	Next, the optimal payoff for this case is given by
	\begin{equation}
	H_{r_{i}=0}^{*}=\begin{cases}
	-\frac{\hat{b_{i}}^2\gamma_{i}}{2}+b_{i}\hat{b_{i}}\gamma_{i} & 0 \leq \hat{b_{i}} \leq b_{i}+p/\gamma_{i} \\
	\frac{p^{2}}{2\gamma_{i}}+\frac{\gamma_{i}b_{i}^{2}}{2}+p(b_{i}-\hat{b_{i}}) & b_{i}+p/\gamma_{i} \leq \hat{b_{i}} \leq q_{max,i}  
	\end{cases} \label{A1}
	\end{equation}

Then the same condition but the utility function is saturated. 	
	
	b) $r_{i}=0$, $\hat{b_{i}}\geq q_{i}$, $q_{i}>b_{i}+\frac{p}{\gamma_{i}}$, $q_{i}\leq q_{max,i}$, $0 \leq \hat{b_{i}} \leq q_{max,i}$
	\begin{alignat*}{3}
	&q_{i,r_{i}=0}^{*}=\quad&& \argmax_{q_{i,r_{i}=0}} \quad && H_{r_{i}=0}=\frac{p^{2}}{2\gamma_{i}}+\frac{\gamma_{i}b_{i}^{2}}{2}+pb_{i}-p\hat{b_{i}}\\
	& && \text{s.t.}&& -\hat{b_{i}}+q_{i}\leq 0 \\
	& && && -q_{i}+b_{i}+p/\gamma_{i} \leq 0 \\
	& && && q_{i}-q_{max,i} \leq 0
	\end{alignat*}
	The first-order optimality condition yields to
	\[
	q_{i,r_{i}=0}^{*}\in [b_{i}+p/\gamma_{i},q_{max,i}],\quad b_{i}+p/\gamma_{i} \leq \hat{b_{i}} \leq q_{max,i}
	\]
	Later, the optimal payoff for this case results to
	\begin{equation}
	H_{r_{i}=0}^{*}=\frac{p^{2}}{2\gamma_{i}}+\frac{\gamma_{i}b_{i}^{2}}{2}+p(b_{i}-\hat{b_{i}}),\quad b_{i}+p/\gamma_{i} \leq \hat{b_{i}} \leq q_{max,i} \label{A2}
	\end{equation}
	
The case c) is assumed when reported baseline is lower than the actual consumption, and the utility function is non-saturated part, i.e.,	
	
	c) $r_{i}=0$, $\hat{b_{i}} \leq q_{i}$, $0\leq q_{i}\leq b_{i}+\frac{p}{\gamma_{i}}$, $0 \leq \hat{b_{i}} \leq q_{max,i}$
	\begin{alignat*}{3}
	&q_{i,r_{i}=0}^{*}=\quad&& \argmax_{q_{i,r_{i}=0}} \quad && H_{r_{i}=0}=-\frac{\gamma_{i}}{2}q_{i}^{2}+[\gamma_{i}b_{i}+p]q_{i}-pq_{i}\\
	& && \text{s.t.}&& \hat{b_{i}}-q_{i}\leq 0 \\
	& && && q_{i}-b_{i}-p/\gamma_{i} \leq 0 \\
	& && && -q_{i} \leq 0
	\end{alignat*}
	The first-order optimality condition yields to	
	\[
	q_{i,r_{i}=0}^{*}=\begin{cases}
	b_{i}& 0 \leq \hat{b_{i}} \leq b_{i}\\
	\hat{b_{i}} & b_{i} \leq \hat{b_{i}} \leq b_{i}+p/\gamma_{i}
	\end{cases}
	\]
	Next, the optimal payoff for this case is given by
	
	\begin{equation}
	H_{r_{i}=0}^{*}=\begin{cases}
	\frac{\gamma_{i}b_{i}^{2}}{2}& 0 \leq \hat{b_{i}} \leq b_{i}\\
	-\frac{\hat{b_{i}}^2\gamma_{i}}{2}+b_{i}\hat{b_{i}}\gamma_{i} & b_{i} \leq \hat{b_{i}} \leq b_{i}+p/\gamma_{i} 
	\end{cases} \label{A3}
	\end{equation}

Finally, the last case is when reported baseline is lower than the actual consumption and the utility function is saturated.

	
	d) $r_{i}=0$, $\hat{b_{i}} \leq q_{i}$, $q_{i}>b_{i}+\frac{p}{\gamma_{i}}$, $q_{i}\leq q_{max,i}$, $0 \leq \hat{b_{i}} \leq q_{max,i}$
	\begin{alignat*}{3}
	&q_{i,r_{i}=0}^{*}=\quad&& \argmax_{q_{i,r_{i}=0}} \quad && H_{r_{i}=0}=\frac{p^{2}}{2\gamma_{i}}+\frac{\gamma_{i}b_{i}^{2}}{2}+pb_{i}-pq_{i}\\
	& && \text{s.t.}&& \hat{b_{i}}-q_{i}\leq 0 \\
	& && && -q_{i}+b_{i}+p/\gamma_{i} \leq 0 \\
	& && && q_{i}-q_{max,i} \leq 0
	\end{alignat*}
	The first-order optimality condition yields to	
	
	\[
	q_{i,r_{i}=0}^{*}=b_{i}+p/\gamma_{i}, \quad 0 \leq \hat{b_{i}} \leq b_{i}+p/\gamma_{i} 
	\]
	And the optimal payoff results to 	
	
	\begin{equation}
	H_{r_{i}=0}^{*}=-\frac{p^{2}}{2\gamma_{i}}+\frac{\gamma_{i}b_{i}^{2}}{2}, \quad 0 \leq \hat{b_{i}} \leq b_{i}+p/\gamma_{i} \label{A4}
	\end{equation}

	\begin{figure}[ht]
		\begin{center}
			\includegraphics[scale=1]{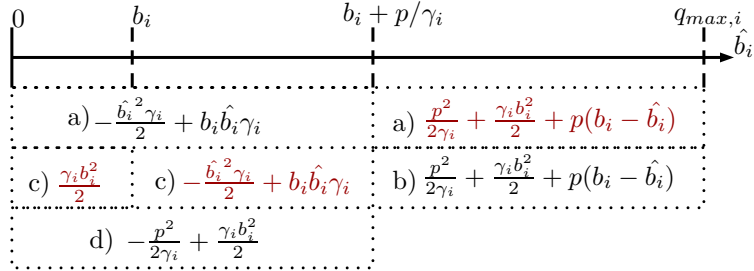}
		\end{center}
		\caption{ \label{fig:comparisionr0} Comparison of optimal profits at Stage 2 when $r_i=0$.}
	\end{figure}
	
Fig. \ref{fig:comparisionr0} shows the comparison of optimal payoff (\ref{A1}), (\ref{A2}), (\ref{A3}) and (\ref{A4}). The final solution is found by selecting the maximum profit according to the reported baseline. Red expressions in Fig. \ref{fig:comparisionr0} corresponds to maximum values.      	
	
	\begin{figure}[ht]
		\begin{center}
			\includegraphics[scale=1.2]{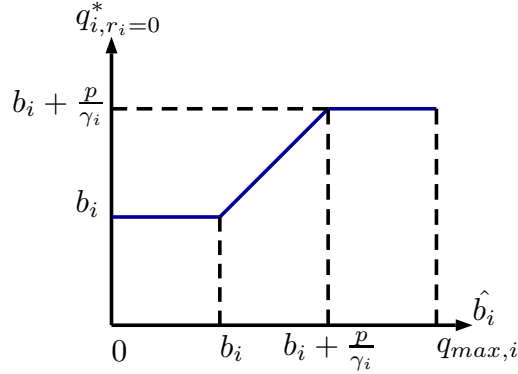}
		\end{center}
		\caption{\label{fig:solr0}Optimal solution at Stage 1 for $r_i=0$.}
	\end{figure}

Finally, the optimal decision at the Stage 1 is drawn in Fig. \ref{fig:solr0}. The result is given by Theorem \ref*{thr0}.

\end{proof}

\section{Proof of the Theorem \ref{thr1}}\label{ath1}

\begin{proof}
	The procedure to solve the Stage 2 when $r_i=1$ is similar to Theorem \ref{thr0}. Although, the results are given by the reported baseline and the reduced energy consumption under DR. The optimization problem is written as follows:  
	\[
	q_{i,r_{i}=1}^{*}=\argmax_{q_{i,r_{i}=1}\in\{0, q_{max,i}\}}\:\, H_{r_{i}=1}=G(q_{i,r_{i}};b_{i})-[pq_{i}-p_{2}(\hat{b_{i}}-q_{i})_{+}+p_{3}\left|q_{i}-\hat{q_{i}}\right|]
	\]
	Since the problem is nonlinear then eight cases are identified but six of them are feasible due to Assumption \ref{asum:logical}. Two parts of utility function $G(.)$, two combinations for incentive expression, and two regions in the penalty of deviation are taken into account in this analysis. All cases and their solutions are listed below.

	e) $r_{i}=1$, $q_{i} \neq \hat{q_{i}}$, $0 \leq q_{i} \leq b_{i}+p/\gamma_{i}$, $q_{i}\geq \hat{b_{i}}$, $q_{i}-\hat{q_{i}} > 0$,  $0 \leq \hat{b_{i}} \leq q_{max,i}$,  $0 \leq \hat{q_{i}} \leq q_{max,i}$

	\[
	q_{i,r_{i}=1}^{*}=\begin{cases}
	(b_{i}-\frac{p_2}{\gamma_{i}})_{+}& 0 \leq \hat{b_{i}} \leq b_{i}-\frac{p_{2}}{\gamma_{i}}, 0 \leq \hat{q_{i}} \leq b_{i}-\frac{p_{2}}{\gamma_{i}} \qquad e)\,1\\
	\hat{b_{i}}& b_{i}-\frac{p_{2}}{\gamma_{i}} \leq \hat{b_{i}}\leq b_{i}+\frac{p}{\gamma_{i}}, 0 \leq \hat{q_{i}} \leq \hat{b_{i}}\qquad e)\,2  \\
	\end{cases}
	\]

	\begin{equation}
	H_{r_{i}=1}^{*}=\begin{cases}
\frac{b_{i}^2\gamma_{i}}{2}-b_{i}p_{2}+\frac{p_{2}^2}{2\gamma_{i}}+\hat{q_{i}}p_{2}& 0 \leq \hat{b_{i}} \leq b_{i}-\frac{p_{2}}{\gamma_{i}}, 0 \leq \hat{q_{i}} \leq b_{i}-\frac{p_{2}}{\gamma_{i}}\qquad e)\,1\\
p_{2}\hat{q_{i}}-\hat{b_{i}}p_{2}-\frac{\hat{b_{i}}^2\gamma_{i}}{2}+b_{i}\hat{b_{i}}\gamma_{i}&  b_{i}-\frac{p_{2}}{\gamma_{i}} \leq \hat{b_{i}}\leq b_{i}+\frac{p}{\gamma_{i}}, 0 \leq \hat{q_{i}} \leq \hat{b_{i}}\qquad e)\,2  \\
\end{cases} \label{A5}
	\end{equation}

	f) $r_{i}=1$, $q_{i} \neq \hat{q_{i}}$, $0 \leq q_{i} \leq b_{i}+p/\gamma_{i}$, $q_{i}\leq \hat{b_{i}}$, $q_{i}-\hat{q_{i}} > 0$, $0 \leq \hat{b_{i}} \leq q_{max,i}$,  $0 \leq \hat{q_{i}} \leq q_{max,i}$

	\[
	q_{i,r_{i}=1}^{*}=\begin{cases}
	(b_{i}-\frac{2p_{2}}{\gamma_{i}})_{+}& b_{i}-\frac{2p_2}{\gamma_{i}} \leq \hat{b_{i}} \leq q_{max,i} , 0 \leq \hat{q_{i}} \leq b_i-\frac{2p_2}{\gamma_{i}} \qquad f)\,1   \\
	\hat{b_{i}}& 0 \leq \hat{b_{i}} \leq b_{i}-\frac{2p_2}{\gamma_{i}} , 0 \leq \hat{q_{i}} \leq \hat{b_{i}} \leq  b_{i}-\frac{2p_2}{\gamma_{i}} \qquad f)\,2  \\
	\hat{q_{i}} & 0 \leq \hat{q_{i}} \leq \hat{b_{i}} \leq q_{max,i}, b_i - \frac{2p_2}{\gamma_{i}} \leq \hat{q_{i}} \leq b_i+\frac{p}{\gamma_{i}}\qquad f)\,3 \\
	\end{cases}
	\]
	
{\footnotesize
	\begin{equation}
		H_{r_{i}=1}^{*}=\begin{cases}
	\frac{2p_2^2}{\gamma_{i}}-2b_ip_2+\hat{b_{i}}p_2+p_2\hat{q_{i}}+\frac{b_i^2\gamma_{i}}{2}& b_{i}-\frac{2p_2}{\gamma_{i}} \leq \hat{b_{i}} \leq q_{max,i} , 0 \leq \hat{q_{i}} \leq b_i-\frac{2p_2}{\gamma_{i}} \qquad f)\,1 \\
	p_{2}\hat{q_{i}}-\hat{b_{i}}p_{2}-\frac{\hat{b_{i}}^2\gamma_{i}}{2}+b_{i}\hat{b_{i}}\gamma_{i}&  0 \leq \hat{b_{i}} \leq b_{i}-\frac{2p_2}{\gamma_{i}} , 0 \leq \hat{q_{i}} \leq \hat{b_{i}} \leq  b_{i}-\frac{2p_2}{\gamma_{i}} \qquad f)\,2  \\
	\hat{b_{i}}p_{2}-p_{2}\hat{q_{i}}-\frac{\hat{q_{i}}^2\gamma_{i}}{2}+b_{i}\hat{q_{i}}\gamma_{i} &  0 \leq \hat{q_{i}} \leq \hat{b_{i}} \leq q_{max,i}, b_i - \frac{2p_2}{\gamma_{i}} \leq \hat{q_{i}} \leq b_i+\frac{p}{\gamma_{i}} \qquad f)\,3 \\
	\end{cases} \label{A6}
	\end{equation}
	}
	
	g) $r_{i}=1$, $q_{i} \neq \hat{q_{i}}$, $q_{i} \geq b_{i} +p/\gamma_{i}$, $q_{i}\geq \hat{b_{i}}$, $q_{i}\leq q_{max,i}$, $q_{i}-\hat{q_{i}} > 0$, $0 \leq \hat{b_{i}} \leq q_{max,i}$,  $0 \leq \hat{q_{i}} \leq q_{max,i}$

	\[
	q_{i,r_{i}=1}^{*}=b_{i}+p/\gamma_{i} \quad 0 \leq \hat{b_{i}} \leq b_{i}+p/\gamma_{i}, 0 \leq \hat{q_{i}} \leq b_{i}+p/\gamma_{i}
	\]

	\begin{equation}
		H_{r_{i}=1}^{*}=\frac{\gamma_{i}b_{i}^2}{2}-p_2b_{i}-\frac{p^2}{2\gamma_{i}}-\frac{p_2p}{\gamma_{i}}+p_2\hat{q_{i}} \label{A7}
	\end{equation}

	h) $r_{i}=1$, $q_{i} \neq \hat{q_{i}}$, $q_{i} \geq b_{i} +p/\gamma_{i}$, $q_{i}\leq \hat{b_{i}}$, $q_{i}\leq q_{max,i}$, $q_{i}-\hat{q_{i}} > 0$, $0 \leq \hat{b_{i}} \leq q_{max,i}$,  $0 \leq \hat{q_{i}} \leq q_{max,i}$

	\[
	q_{i,r_{i}=1}^{*}=b_{i}+p/\gamma_{i} \quad b_{i}+p/\gamma_{i}\leq \hat{b_{i}} \leq q_{max,i}  , 0 \leq \hat{q_{i}} \leq b_{i}+p/\gamma_{i}
	\]
	
	\begin{equation}
	H_{r_{i}=1}^{*}=\frac{\gamma_{i}b_{i}^2}{2}-2p_{2}b_{i}-\frac{p^2}{2\gamma_{i}}-\frac{2p_{2}p}{\gamma_{i}}+\hat{b_{i}}p_{2}+p_{2}\hat{q_{i}} \label{A8}
	\end{equation}

	i) $r_{i}=1$, $q_{i} \neq \hat{q_{i}}$, $0 \leq q_{i} \leq b_{i}+p/\gamma_{i}$, $q_{i}\geq \hat{b_{i}}$, $q_{i}-\hat{q_{i}} < 0$, $0 \leq \hat{b_{i}} \leq q_{max,i}$,  $0 \leq \hat{q_{i}} \leq q_{max,i}$. 	Case i) is infeasible by Assumption \ref{asum:logical}.

	j) $r_{i}=1$, $q_{i} \neq \hat{q_{i}}$, $0 \leq q_{i} \leq b_{i}+p/\gamma_{i}$, $q_{i}\leq \hat{b_{i}}$, $q_{i}-\hat{q_{i}} < 0$, $0 \leq \hat{b_{i}} \leq q_{max,i}$,  $0 \leq \hat{q_{i}} \leq q_{max,i}$

	\[
	q_{i,r_{i}=1}^{*}=\begin{cases}
	b_{i}& b_{i} \leq \hat{b_{i}} \leq q_{max,i}, b_{i} \leq \hat{q_{i}} \leq q_{max,i} \qquad j)\,1\\
	\hat{q_{i}}& \hat{q_{i}} \leq \hat{b_{i}} \leq q_{max,i},  0 \leq \hat{q_{i}} \leq b_{i}  \qquad j)\,2\\
	\end{cases}
	\]

	\begin{equation}
H_{r_{i}=1}^{*}=\begin{cases}
\hat{b_{i}}p_{2}-p_{2}\hat{q_{i}}+\frac{b_{i}^2\gamma_{i}}{2}  & b_{i} \leq \hat{b_{i}} \leq q_{max,i}, b_{i} \leq \hat{q_{i}} \leq q_{max,i} \qquad j)\,1 \\
\hat{b_{i}}p_{2}-p_{2}\hat{q_{i}}-\frac{\hat{q_{i}}^2\gamma_{i}}{2}+b_i\hat{q_{i}}\gamma_{i}  &\hat{q_{i}} \leq \hat{b_{i}} \leq q_{max,i},  0 \leq \hat{q_{i}} \leq b_{i}  \qquad j)\,2 \\
\end{cases} \label{A9}
	\end{equation}

	k) $r_{i}=1$, $q_{i} \neq \hat{q_{i}}$, $q_{i} \geq b_{i} +p/\gamma_{i}$, $q_{i}\geq \hat{b_{i}}$, $q_{i}\leq q_{max,i}$, $q_{i}-\hat{q_{i}} < 0$, $0 \leq \hat{b_{i}} \leq q_{max,i}$,  $0 \leq \hat{q_{i}} \leq q_{max,i}$. As well, case k) is infeasible.	
	
	Finally, case l) is solved.	
	
	l) $r_{i}=1$, $q_{i} \neq \hat{q_{i}}$, $q_{i} \geq b_{i} +p/\gamma_{i}$, $q_{i}\leq \hat{b_{i}}$, $q_{i}\leq q_{max,i}$, $q_{i}-\hat{q_{i}} < 0$, $0 \leq \hat{b_{i}} \leq q_{max,i}$,  $0 \leq \hat{q_{i}} \leq q_{max,i}$.

	\[
	q_{i,r_{i}=1}^{*}=b_{i}+p/\gamma_{i} \quad b_{i}+p/\gamma_{i} \leq \hat{b_{i}} \leq q_{max,i} , b_{i}+p/\gamma_{i} \leq \hat{q_{i}} \leq q_{max,i}
	\]
	
	\begin{equation}
	H_{r_{i}=1}^{*}=\frac{\gamma_{i}b_i^2}{2}-\frac{p^2}{2\gamma_{i}}+\hat{b_{i}}p_2-p_2\hat{q_{i}} \label{A10}
	\end{equation}

	\begin{figure}[ht]
		\begin{center}
			\includegraphics[scale=0.6]{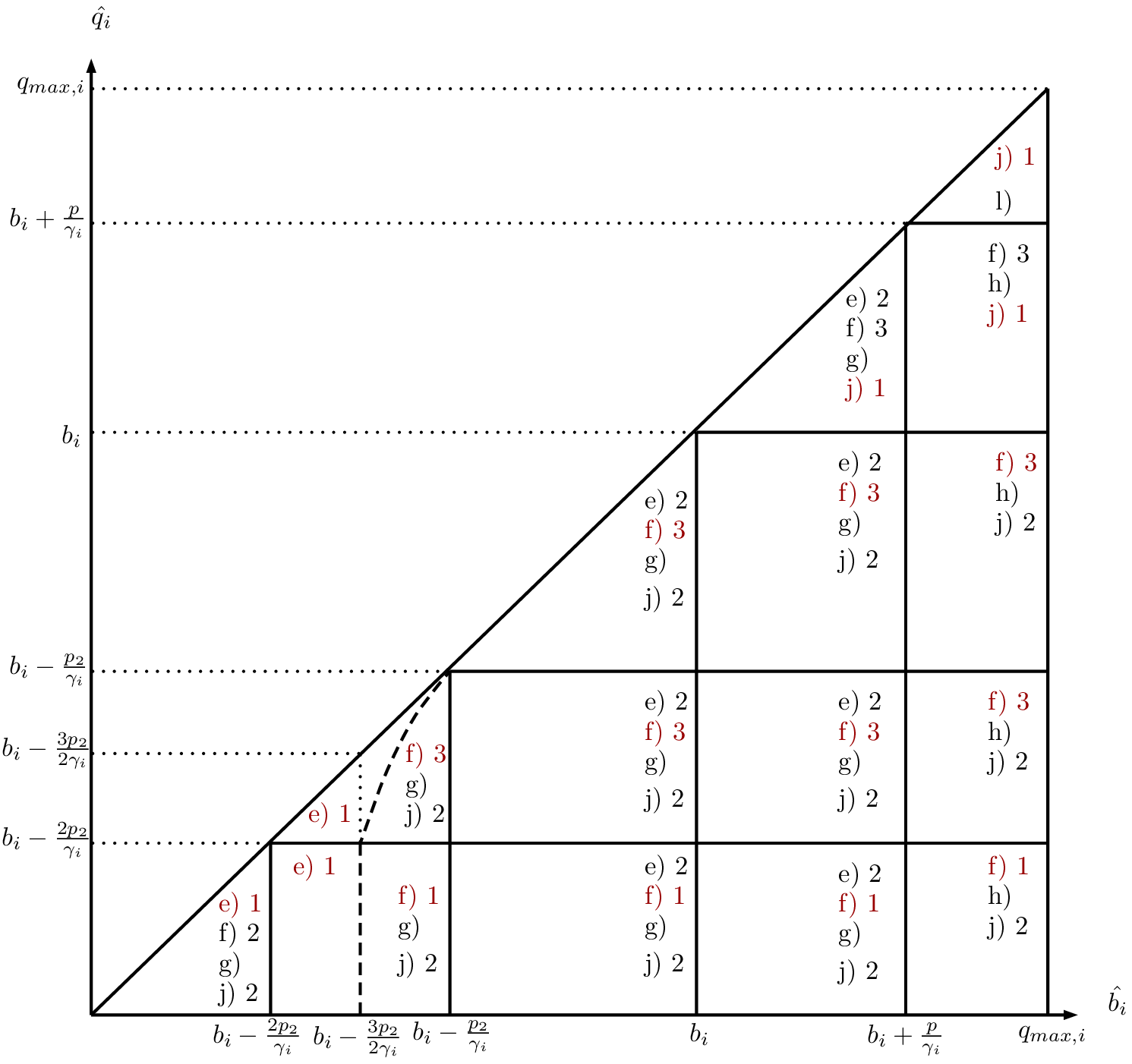}
		\end{center}
		\caption{\label{fig:comparisionr1} Comparison of optimal profits at Stage 2 when $r_i=1$.}
	\end{figure}

	Next, there are two local maxima when solution e) 1 and f) 1 are found in the same feasible region. In order to find the global solution, the payoff in cases e) 1 and f) 1 are compared. The critical value of $\hat{b_{i}}$ that provides the same payoff in both cases is:

	$\frac{b_{i}^2\gamma_{i}}{2}-b_{i}p_{2}+\frac{p_{2}^2}{2\gamma_{i}}+\hat{q_{i}}p_{2}=\frac{2p_2^2}{\gamma_{i}}-2b_ip_2+\hat{b_{i}}p_2+p_2\hat{q_{i}}+\frac{b_i^2\gamma_{i}}{2}$, 
	
	Then solving for $\hat{b_{i}}$,
	
	$\hat{b_{i}}=b_i-\frac{3p_2}{2\gamma_{i}}$ 
	
	Similarly, the same situation occurs for e) 1 and f) 3. The critical value is renamed as $\alpha$ because depends on reported baseline $\hat{b_{i}}$ and the announce of reduced energy consumption $\hat{q_{i}}$. This value is found in Theorem \ref{thr1}. 
	
	Fig. \ref{fig:comparisionr1} presents the comparison of optimal payoff (\ref{A5}), (\ref{A6}), (\ref{A7}), (\ref{A8}), (\ref{A9}) and (\ref{A10}). The final solution is found by choosing the maximum profit accordig to $\hat{b_{i}}$ and $\hat{q_{i}}$. Red expression in Fig. \ref{fig:comparisionr1} corresponds to maximum values. The result is given by Theorem \ref{thr1}.  
\end{proof}

\section{Proof of the Theorem \ref{threport}}\label{athreport}

\begin{proof}
	
The results from Theorems \ref*{thr0} and \ref*{thr1} are substituted in Eq. (\ref{stage1}) in order to solve the optimization problem. Fig. \ref{fig:intersection} presents the intersection of strategies at Stage 2. Every crossing between results when $r_i=0$ and $r_i=1$ in (\ref{stage1}) is solved. Next, the outcomes are compared according to the reported baseline and reduced energy consumption in order to determine the global maximum. The final result is given by Theorem \ref{threport}.

	\begin{figure}[ht]
		\begin{center}
			\includegraphics[scale=0.6]{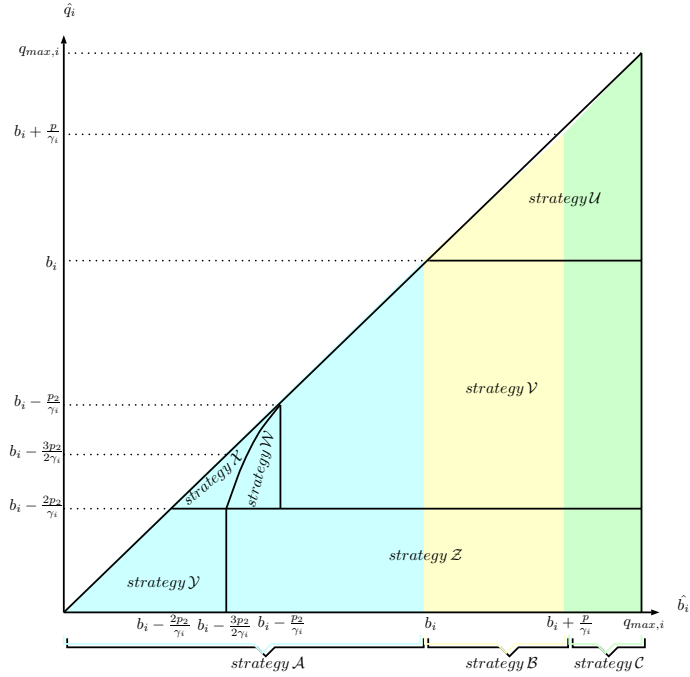}
		\end{center}
		\caption{ \label{fig:intersection}Intersection of feasible regions.}
	\end{figure}
\end{proof}

\bibliographystyle{elsarticle-harv}
\bibliography{mechanism}

\end{document}